# A New Citation Recommendation Strategy Based on Term Functions in Related Studies Section

Haihua Chen†

Department of Information Science, University of North Texas, Texas 76207, USA

**Abstract**

**Purpose:** Researchers frequently encounter the following problems when writing scientific articles: (1) Selecting appropriate citations to support the research idea is challenging. (2) The literature review is not conducted extensively, which leads to working on a research problem that others have well addressed. This study focuses on citation recommendation in the related studies section by applying the term function of a citation context, potentially improving the efficiency of writing a literature review.

**Design/methodology/approach:** We present nine term functions with three newly created and six identified from existing literature. Using these term functions as labels, we annotate 531 research papers in three topics to evaluate our proposed recommendation strategy. BM25 and Word2vec with VSM are implemented as the baseline models for the recommendation. Then the term function information is applied to enhance the performance.

**Findings:** The experiments show that the term function-based methods outperform the baseline methods regarding the recall, precision, and F1-score measurement, demonstrating that term functions are useful in identifying valuable citations.

**Research limitations:** The dataset is insufficient due to the complexity of annotating citation functions for paragraphs in the related studies section. More recent deep learning models should be performed to future validate the proposed approach.

**Practical implications:** The citation recommendation strategy can be helpful for valuable citation discovery, semantic scientific retrieval, and automatic literature review generation.

**Originality/value:** The proposed citation function-based citation recommendation can generate intuitive explanations of the results for users, improving the transparency, persuasiveness, and effectiveness of recommender systems.

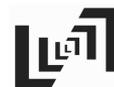

† Corresponding author: Haihua Chen (Email: Haihua.Chen@unt.edu).

## 1　Introduction

The amount of scientific literature has been increasing exponentially in recent years. For example, publications in Computer Science in Web of Science have been growing from 396 in 1995 to 50,644 in 2021[①]. Due to the explosion of scientific literature, it has become more and more time-consuming for readers to review related literature and decide which article to cite. Fortunately, citation recommendation (CR) has been proved to be useful in helping users to decide which papers should be cited from their reading list (Beel & Dinesh, 2017; He et al., 2010; Liu, Yan, & Yan, 2013; McNee et al., 2002; Raamkumar, Foo, & Pang, 2016; Strohman, Croft, & Jensen, 2007). A CR system suggests previous studies be reviewed and cited for new research articles. It helps researchers to cite appropriate previous studies and to avoid missing important literature. Usually, an automatic CR system accepts a research topic and provides a list of publications that can be cited. Unlike traditional search approaches offered by search engines and digital libraries, CR systems focus on finding the relevant publications rather than any texts or pages similar to the topics (Strohman, Croft, & Jensen, 2007).

Research has applied semantic information and non-semantic information for CR. Semantic information such as citation function, citation sentiment, citation importance, term function, is usually included in the citation context, which is defined as a sequence of words appearing around a citation placeholder. For example, He et al. (2010) designed a non-parametric probabilistic CR model which measured the context-based relevance between a citation context and a document to be recommended and automatically identified citation contexts in a manuscript where citations were needed by applying contextual information (He et al., 2011). Tang and Zhang (2009) discovered topical aspects of the citation contexts of each paper and recommended papers based on the discovered topic distribution. Zarrinkalam and Kahani (2013) used citation context as a textual feature to enhance the performance of CR tasks. Duma et al. (2016a) and Duma (2019) integrated core scientific concepts classification and discourse annotation into context-based CR.

Non-semantic information mainly focuses on the relationship between articles and authors, such as citation network, author collaboration, co-occurrence. For example, McNee et al. (2002) created a rating matrix using the citation network between papers. Chen et al. (2011) also leveraged such citation-network-based methodology but named it citation authority diffusion (CAD). Livne et al. (2014) developed a CR system that took the author similarity, venue relevancy, and

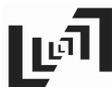

---

[①] https://apps.webofknowledge.com/UA_GeneralSearch_input.do?product=UA&SID=7F4FLVerSQulGPpm4AT&search_mode=GeneralSearch

co-citation into consideration for augmenting sparse citation networks. Son and Kim (2018) proposed a multilevel citation network-based scientific paper recommender system by comparing all the indirectly linked papers to the paper of interest. It can recommend both the research topic and the academic theory related papers. Some scholars also combined semantic information with non-semantic information to enhance the CR performance (Bethard & Jurafsky, 2010; Ebesu & Fang, 2017; Jeong et al. 2019; Strohman, Croft, & Jensen, 2007).

Current CR systems could be further improved by considering new factors. One promising approach to enhance CR service is to recommend citations in the related studies sections. Citation content analysis results showed that more than 60% of the references and the most highly cited articles appeared in the introduction and literature review sections (collectively referred to as related studies sections in this paper) of the citing papers (Ding et al., 2013). Recommending citations in related work sections has been discussed as meaningful to fascinating literature review writing (Huang et al., 2014; Livne et al., 2014). However, little research has been conducted on this valuable task, except for (Sesagiri Raamkumar, Foo, & Pang, 2015; 2016), who constructed a recommender system for providing a shortlist of papers based on article type preference, coverage, citation count, and user-specified prospective keywords to assist researchers' literature review writing. The drawback is that users still need to spend much energy on how to organize these articles. We believe that a CR algorithm which can recommend papers categorized by their term functions in related studies sections will save users a lot of time.

In this paper, the term function refers to the semantic role of a segment or a paragraph in the related studies section (Cheng, 2015). Term function has been proved useful for academic retrieval and recommendation (Li, Cheng, & Lu, 2017). For example, when people conduct research on citation context recognition (CCR), the related work may involve problem statement on CCR, the CCR methods, datasets used in CCR, CCR related tools, CCR evaluation method, and the applications of CCR. Therefore, citation recommendation services can provide recommendation lists according to these "term functions" based on users' requirements, this will be more likely to meet their information needs. This paper focuses on this innovative and challenging task, aiming to explore a more efficient CR strategy thus improving scholars' reading experience. Compared with previous studies, the contributions of this paper mainly include three aspects:

- We investigate the citation organization patterns in the related work sections. Scientific articles tend to follow a particular style of organizing sections and paragraphs (Luong, Nguyen, & Kan, 2012). To understand how researchers usually organize related studies, we develop a term function annotation scheme

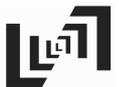

- at the paragraph-level. An annotation experiment showed that there were four common patterns of organizing literature in the related work sections.
- We propose a term function-based citation recommendation framework to recommend articles for users based on their expected term function of a specific paragraph in the related work sections. This is the first framework to introduce term function into citation recommendation task to the best of our knowledge.
- We conduct several experiments on "real-world" datasets obtained from ACL Anthology to evaluate the impacts of the term function factors and the performance of the proposed methods. The experimental results show that our proposed Word2vec-VSM model with term functions achieve the best recommendation performance, indicating the effectiveness of term functions on citation recommendation.

The remainder of this paper is structured as follows: Section 2 reviews the related work. Section 3 introduces the proposed framework for term function-based citation recommendation, including the term function annotation experiment and the recommendation algorithms. Section 4 describes our recommendation experiments based on the ACL Anthology dataset and reports the results. Section 5 concludes the paper and discusses the future study.

## 2. Related work

In this study, we assume that paragraphs in related work sections could be organized by the term function of the citation sentences in them. Therefore, our work in this paper first annotates each paragraph in related work sections with a particular term function, then combines with paragraph content and term function weight to recommend citations for this paragraph. Given this focus, we review scientific literature related to term function and citation recommendation.

### 2.1 Term function

Initially, term function refers to the semantic role that a term plays in the scientific literature (Cheng, 2015). It also represents the sentence's function where the term belongs if all the citation sentences in a paragraph in the related work section share a specific "term function" (for example, the research method related to the citing article). We believe this paragraph will be easier to understand because all the citations focus on a specific term function (of a topic).

Except for the term function "research method" we mentioned above, there are many other term functions, such as research topic, technology dataset, application, evaluation. However, according to previous definitions, term functions could be classified into different schemes, as shown in table 1.

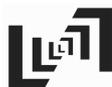

Table 1.   Classification of term function (Li, Cheng, & Lu, 2017).

| Classification of term function | Authors |
| --- | --- |
| Head, goal, method, other | Kondo (2009) |
| Technology, effect | Nanba, Kondo, & Takezawa (2010) |
| Focus, technique, domain | Gupta & Manning (2012) |
| Technique, application | Tsai, Kundu, & Roth (2013) |
| Method, task, other | Huang & Wan (2013) |
| Domain-independent: Research topic, Research method | Cheng (2015) |
| Domain-related: Case, tool, dataset, etc. | |

However, the classification of the term function is still obscure and not uniform. Considering the specific research problem in this article, we took some categories from existing literature and proposed three new categories of term function, which we will introduce in the next section. As a novel task, although several attempts have been made on term function analysis, the topic still remains to be developed, especially the gap between term function analysis and its applications, which forms the exact initiation of our study.

## 2.2   Citation recommendation

Citation recommendation for research papers was first introduced by McNee et al. (2002). Since then, there has been a rich line of research on this topic. For example, Strohman, Croft, and Jensen (2007) combined the content of previous literature and its citation network to recommend relevant material that a given article should cite. They found that this mixed method performed much better than the traditional text-similarity approach. Tang and Zhang (2009) proposed a topic distribution discovery-based CR model that performed well on sentence-level CR on the NIPS dataset. They argued that CR could be integrated into academic search systems to improve service. He et al. (2011) presented a CR system that automatically identified contexts where citations were needed in a manuscript and recommended appropriate citations. In 2013, Kates-harbeck and Haggblade (2013) used a machine learning method that utilized context-based features and text-based features to rank references for a given short text. Recently, with the growth of scholarly data and the development of deep learning, neural networks were transplanted to CR problems, aiming to train more robust models and enhance CR performance (Ebesu & Fang, 2017; Huang et al., 2015).

In terms of the context scope based on which a citation recommendation list was generated, the CR task can be divided into two aspects: local citation recommendation (LCR) (Tang, Wan, & Zhang, 2014; Yang et al., 2019) and global citation recommendation (GCR) (Tang, Wan, & Zhang, 2014; Ayala-Gomez et al., 2018).

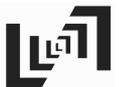

### 2.2.1 Local citation recommendation

Local citation recommendation aims to recommend citations for a specific context where citations are needed, which is also called context-aware citation recommendation (Tang, Wan, & Zhang, 2014). Here, the specific context can be one sentence or several sentences; in this paper, it represents all the sentences in a paragraph in the related work sections. Since contexts contain rich semantic information, content-based approaches were usually used in LCR.

He et al. (2010) proposed an effective context-aware citation recommendation approach, which designed a non-parametric probabilistic model to measure the relevance between a citation context and a document. In a similar task, the translation model was used to translate the query terms in the citation context, which bridged the vocabulary gap between the citation context and recommended document (Lu et al., 2011; He et al., 2012). Rokach et al. (2013) presented a supervised learning method utilizing three types of features (general features, author-aware features, and context-aware features) to recommend citations for a given citation context. This approach had been applied for citation recommendation service in CiteSeerX digital library. Except for these features, time or publication date has also been proven to be important in citation recommendation (Gao, 2016; Jiang, 2015; Jiang, Liu, & Gao, 2014; 2015). To provide a personalized citation recommendation service, Liu, Yan, and Yan (2013) combined user profiles with context-aware citation recommendation. Experimental results showed that the proposed strategy outperformed language model-based and translation model-based algorithms. Other research explored the effectiveness of citation networks (Jiang et al., 2018; Livne et al., 2014; Son and Kim, 2018), core scientific concepts (Duma et al., 2016) in local citation recommendation. (Duma & Klein, 2014) introduced a citation resolution method to evaluate context-based citation recommendation systems. Citation context, as the primary evidence to recommend citations, has become a theorem in this field (Bhagavatula et al., 2018).

Recently, with the popularity of deep learning in both academia and industry, different deep learning algorithms have been applied to citation recommendation. (Huang et al., 2014) represented words and documents by learning simultaneously from citation context and cited document pairs; the probabilistic neural model was used to estimate the probability of words appeared in the citation context under a candidate reference paper. This approach achieved 5% improvement on Recall than the translation model. Ebesu and Fang (2017) proposed a neural citation network (NCN) that can model the semantic composition of citation contexts and corresponding cited documents titles by exploiting author relations. This method improved several state-of-the-art baselines on all metrics (Recall, MAP, MRR, and NDCG) by 13–16%.

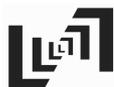

#### 2.2.2 Global citation recommendation

Unlike LCR, global citation recommendation targets on recommending a reference list for a given paper (Tang, Wan, & Zhang, 2014). However, only focus on finding the relevant papers rather than the essential papers can be a drawback of GCR because users need to evaluate the quality of the recommended papers (Chen et al. 2011). This is challenge for new researchers.

To bridge this gap, Küçüktunç et al. (2012; 2013; 2015) recommended a diverse reference list which allowed the users to reach either old, well-cited, well-known research papers or recent, less known ones. To improve users' experience, Wu et al. (2012) proposed to recommend references based on their information needs, such as publication time preference, self-citation preference, co-citation preference, and publication reputation preference. However, most existing research used a single data source for the recommendation, which failed to meet users' information needs about different aspects in writing (Zarrinkalam & Kahani, 2012). Therefore, Zarrinkalam and Kahani (2012) introduced a strategy to recommend citations based on multiple linked data sources, which enriched the background data of recommender systems thus enhancing the recommendations.

Kates-harbeck and Haggblade (2013) recommended references for a given abstract with a set of key technical words, an author list, and publication data using a two-stage methodology. In the first stage, they trained a classifier to rank a candidate reference list based on the paper score; in the second stage, they re-ranked the scored candidate papers using connectivity information. Their experiment results showed that text-based features were most effective in rankings. Caragea et al. (2013) compared the performance of the collaborative filtering-based (CF) approach and singular value decomposition-based (SVD) approach on GCR using CiteSeer dataset, finding that SVD performed better than CF since SVD can easily incorporate additional information.

As a benefit of different citation recommendation tasks and algorithms mentioned above, many citation recommendation systems have been developed, for example, ActiveCite (Zhou, 2010), CiteSight (Livne et al., 2014), RefSeer (Huang et al., 2014), ClusCite (Ren et al., 2014), DiSCern (Chakraborty et al., 2015), and Rec4LRW (Sesagiri Raamkumar, Foo, & Pang, 2015). Due to the usefulness but challenge of this task, citation recommendation is attracting more and more attention from researchers.

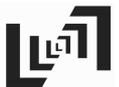

## 3  Proposed approach

The main objective of our work is to recommend citations in the related work sections and verify the effectiveness of term function in this recommendation task. In terms of the first objective, we notice that the literature review sections are usually organized in specific patterns to address different aspects related to an article, thus recommending citations by following these patterns might customize users' personal information needs. As for the second objective, we observe that a specific term in a segment or a paragraph might indicate its role in related work sections; in this paper, we define it as term function, which is valuable information for CR.

The main problems addressed in our work are: (1) identify the paragraph organization patterns in related work sections, and (2) based on these patterns, recommend citations by involving term functions as the weighting parameter. Fig. 1 presents the framework of our proposed approach to solve the problems. Given a user topic, the CR system first retrieves a list of publications as candidates for recommendation. It then assigns weights to the candidates based on their term functions for reranking. At last, the top-ranked candidates are provided to the user.

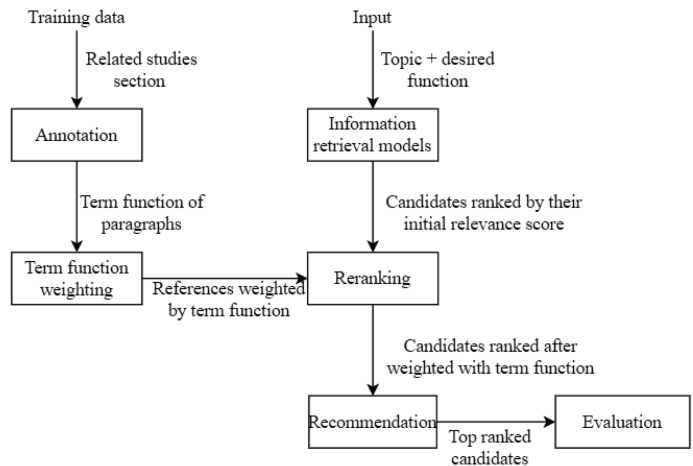

Figure 1.   Framework of term function-based citation recommendation

### 3.1 Paragraph organization patterns analysis in related work sections

Typically, researchers organize related literature by following specific patterns (e.g. based on the similarity, based on different topics, based on the published time, or based on the roles) rather than roughly putting a list of relevant literature together. Meanwhile, the literature is organized to demonstrate previous research problems, methods, datasets, and applications, defined as term functions in our study, that are related to a research. We propose a term function classification scheme used to annotate a real-world dataset acquired from ACL Anthology. After that, we analyze the annotation results to identify the paragraph organization patterns in related work sections.

#### 3.1.1 Term function classification scheme in paragraph level

Term function means the semantic role that a term plays in the scientific literature (Cheng, 2015). The term's function also reflects the function of the citation context it locates in the scientific literature. For example, support vector machine (SVM) is the core research problem in (Li, Wang, & Wang, 2010) but the method to solve image classification problems in (Melgani & Bruzzone, 2004). Therefore, the paper "A Tutorial on Support Vector Machines for Pattern Recognition" (Burges, 1998) is cited as a related research problem in (Li, Wang, & Wang, 2010) but cited as a related method to solve image classification problems in (Melgani & Bruzzone, 2004). In other words, citations could be recommended according to the "term function" needed in the citing article, supposing the related literature in the citing is organized by these term functions. To verify this hypothesis, we develop a term function classification scheme by combining categories from existing literature and three new categories, which we draw from a pilot study, as shown in Table 2. The reason that we add three new functions is existing categories of term functions are proposed for sentence-level. Paragraphs have more complex structures than sentences, and they are usually organized by different roles in the related studies section. For example, a paragraph discusses how the method used in the original paper has been applied to solve other problems; we define this pattern as "Method+ Problem." On the other hand, a paragraph may also focus on discussing how other relevant methods have solved the research problem mentioned in the original paper except for the method proposed in the original paper; we define this pattern as "Problem+ Method." We also define another function, "Topic-irrelevant," since some paragraphs that served as background knowledge are not closely relevant to the original paper's topics. Fig. 3 demonstrates that the three new term functions frequently appear in the related studies section.

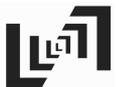

Table 2.   Term functions in the related studies section.

| Category | Source | Description |
| --- | --- | --- |
| Application | Tsai et al. (2013) | Describes existing application of the core problem and method in this article |
| Dataset | Cheng (2015) | Describes related datasets to this article |
| Evaluation | Cheng (2015) | Describes related evaluation methods to this article |
| Method | Huang & Wan (2013) | Describes previous work related to the core method of the article |
| Method+ Problem | **New** | Describes the core method of the article and introduces what problems it can be used to solve |
| Problem | Kondo et al. (2009) | Describes previous work related to the core research problem of the article |
| Problem+ Method | **New** | Describes the core research problem of the article and introduces the existing method to the problem |
| Tool | Cheng (2015) | Describes related tools used in this article |
| Topic-irrelevant | **New** | Describes previous work not very relevant |

### 3.1.2   Annotation

We collected a dataset consisting of 238, 109, and 184 research papers in sentiment analysis, information extraction, and recommender systems, respectively. We manually downloaded the pdf files from ACL Anthology and placed them in three separate folders. We converted the pdf files into text format using Apache PDFBox[②]. We then extracted the title, abstract, and related work section information for annotation.

Paragraph organization pattern annotation is difficult since it requires extensive domain knowledge. To ensure the data quality, we invited two Ph.D. students who are familiar with sentiment analysis, information extraction, and recommender systems to annotate the data. Before annotation, they were required to pre-annotate ten articles under each folder to understand the annotation guideline, shown as follows:

- Firstly, read the title and abstract to get the research problems and research methods of the article.
- Secondly, find the paragraph that needs to be annotated (ignore paragraphs without citations), read the paragraph's contents to get the features that indicate its term function, label the paragraph with a category of term function mentioned above.
- Avoid subjective judgment during annotation.

Fig. 2 describes an annotation example, which was saved into the text format. The pairwise Kappa coefficients (Viera & Garrett, 2005) were applied to calculate

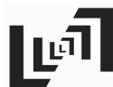

---

[②]   https://pdfbox.apache.org/

the inter-annotator agreement. Cohen's kappa scores were 0.778, 0.871, and 0.748 on sentiment analysis, information extraction, and recommender systems, respectively. These are satisfactory according to the existing evaluation standard (Viera & Garrett, 2005).

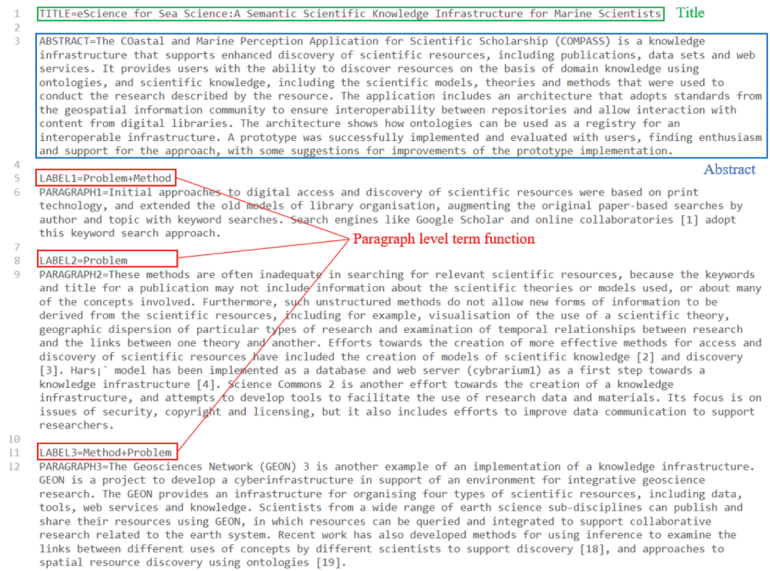

Figure 2.    An annotation example of citation function in paragraph level.

### 3.1.3  Paragraph organization patterns analysis

To investigate how authors organized literature in related work sections, we conducted a statistical analysis of the annotation results. Fig. 3 represents the distribution of term functions under each category over different topics. Based on the results, most paragraphs belong to problem+ method, problem, method+ problem, and method, with a percentage of 38.2%, 29.5%, 15.6%, and 6.0% on average, respectively, but with a slight difference between different topics. The result supports the hypothesis that authors usually pay more attention to the research problems and methods relevant to their topics while writing a literature review. In other words, they tend to focus on investigating who was involved in studying the problems and methods, how they described the problems and methods, what methods have been used to solve the problem, and how the method was applied to other problems. Existing datasets, evaluation methods, tools, and applications related to their research were also occasionally involved in two or more independent paragraphs. In addition, some topic-irrelevant materials were referred to in a few paragraphs, suggesting that the research topic is novel, and existing literature in this area was lacking.

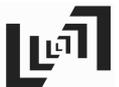

The findings enlighten us to conduct citation recommendation in the related work sections by considering these information needs. In this paper, we explored this innovative citation recommendation task based on four major term functions: problem+ method, problem, method+ problem, and method.

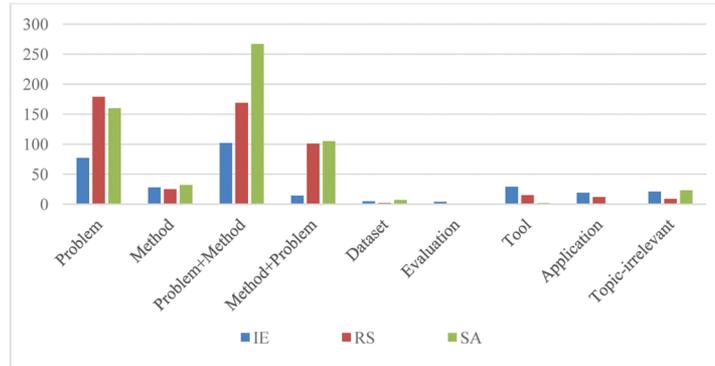

Figure 3.    Statistical analysis of the term function distribution in three fields.

### 3.2    Term function-based citation recommendation

As discussed previously, the term function-based citation recommendation is a kind of global citation recommendation at the paragraph level. Users are required to input the topic (e.g. citation recommendation, translation model) and the term function (e.g. problem+ method, method+ problem). Our approach will recommend a list of citations that are both topic and term function relevant.

#### 3.2.1    Problem definition

**Definition 1** (Original document collection). Given a document $d$ with $t$ and $a$ as its title and abstract, the document has a set of paragraphs $P = \{p_1, p_2, ..., p_i\}$ in the related works section, the term function of each paragraph $p_i$ is defined as $l_i$. Therefore, the paragraph set P has a corresponding term function set $L = \{l_1, l_2, ..., l_i\}$. All the original documents together with their paragraph sets and term function sets form the original document collection $D$.

**Definition 2** (Candidate document collection). Given a document $c$ with $t$ and $a$ as its title and abstract, we define its term function as $F = \{f_1, f_2, f_3, f_4\}$, where $f_1, f_2, f_3, f_4$ refer to problem+ method, problem, method+ problem, and method respectively, which are determined by the term function of its citing paragraphs. All the candidate documents were extracted from the references of the original documents; they form the candidate document collection *C*.

**Definition 3** (Term function-based citation recommendation). Given a paragraph *p* whose topic is *t* and expected term function is *f*, our goal is to estimate the probability of each document in the candidate document collection *C* to be cited in this paragraph.

### 3.2.2 Citation recommendation algorithms

(1) BM25 model

BM25 (Robertson & Zaragoza, 2009) has been widely employed in many information retrieval systems and proved to be effective. Recently, it has also been frequently selected as a baseline model in many citation recommendation tasks (Bhagavatula et al., 2018; Ebesu & Fang 2017; Gao, 2016; Jiang, Liu, & Gao, 2015; Ren et al., 2014; Sesagiri Raamkumar, Foo, & Pang, 2015). In this paper, we use BM25 as the baseline method. With the BM25 ranking function, the relevance score of a candidate document d with respect to a paragraph as a query q is calculated as follows:

$$Score(q,d) = \sum_i^n \log \frac{N - n(q_i) + 0.5}{n(q_i) + 0.5} \cdot \frac{f_i \cdot (k_1 + 1)}{f_i + k_1 \cdot (1 - b + b \cdot \frac{dl}{avgdl})} \quad (1)$$

Where $k_1$ and $b$ are free hyper-parameters. *avgdl* and *N* respectively donate the average document length and the total number of documents in the collection. $f_i$ and $n(q_i)$ represent the frequencies of a query term $q_i$ in a candidate document and the total number of documents containing the query term $q_i$.

(2) Word2vec based vector space model

Word2vec is one of the most widely used word embedding models used to represent texts with numerical vectors (Mikolov et al., 2013). Compared to the term frequency used in the BM25 algorithm, Word2vec can preserve the semantic and syntactic relationship between words. Therefore, we use Word2vec to create vectors for both queries and documents in this paper. We train our Word2vec model with Gensim③, a python package used for topic modeling, text processing, and working with word vector models such as Word2vec and FastText. The Word2vec model is trained on the ACL Anthology Reference Corpus④, which contains 10,920 academic papers from the ACL Anthology. Finally, we generate vectors of size 300 for the

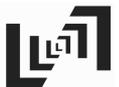

---

③ https://radimrehurek.com/gensim/models/word2vec.html
④ https://web.eecs.umich.edu/~lahiri/acl_arc.html

queries and documents. The similarity between a query and the document is calculated by using the vector space model (VSM) with the following equation:

$$Score(q,d) = \frac{\vec{q} * \vec{d}}{\|\vec{q}\|\|\vec{d}\|} \quad (2)$$

Where $q$ and $d$ are the vector representations of the query and the document, respectively.

### 3.2.3 Term function weighting-based recommendation models

In this paper, we explore the influence of term functions in citation recommendation and propose the term function weighting-based recommendation model. Compared with the original BM25 model and Word2vec-based VSM, an essential step of term function weighting-based recommendation models is to compute the weight of each document in the candidate document collection on each of the four term functions.

We define $F = (f_1, f_2, f_3, f_4)$ as the weighted term function of each document $d$ in the candidate document collection, where $f_1, f_2, f_3, f_4$ respectively represents the frequency of problem, method, problem+ method, method+ problem. After that, the term function should be added to the three recommendation models to improve performance.

For a document $d$ and a paragraph as a query $q$, $q_f$ donates the expected term function of this query/paragraph. Therefore, we calculate the relevance score on the term function dimension as follow:

$$p(q_f, d) = \frac{f}{f_1 + f_2 + f_3 + f_4}, f \in \{f_1, f_2, f_3, f_4\} \quad (3)$$

For example, if a user inputs a query $q$ with the term function as problem, one of a document d in the candidate document collection whose term function is $F = (4, 2, 8, 1)$, we compute its term function weight on problem, method, problem+ method, method+problem respectively as $\frac{4}{15}, \frac{2}{15}, \frac{8}{15}, \frac{1}{15}$, where the sum of the weights should adhere to the following constraint:

$$\sum_{j=1}^{4} weight_j = 1 (weight_j \in [0,1]) \quad (4)$$

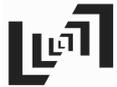

Finally, the relevance score of a candidate document $d$ with respect to a paragraph as a query $q$ which combines text similarity and term function matching is computed with the following two equations respectively:

$$Score(q,d) = (1 + p(q_f, d)) \cdot \sum_{i}^{n} \log \frac{N - n(q_i) + 0.5}{n(q_i) + 0.5} \cdot \frac{f_i \cdot (k_1 + 1)}{f_i + k_1 \cdot (1 - b + b \cdot \frac{dl}{avgdl})} \quad (5)$$

$$Score(q,d) = (1 + p(q_f, d)) \cdot \frac{\bar{q} * \bar{d}}{\|\bar{q}\| \|\bar{d}\|} \qquad (6)$$

Equation (5) is for term function weighting-based BM25 model, while equation (6) is for Word2vec-based VSM models with term function weighting.

Pseudo code for the baseline recommendation models or term function weighting-based recommendation models is described in Algorithm 1.

---

Algorithm 1    Recommendation with baseline models or term function weighting-based recommendation models

---

**Input:**
    Candidate paper list
    Rank papers by relevance scores
    **If** Year of candidate paper < year of original paper
        Add candidate paper into new recommendation list
    **Else**
        Continue
    **End if** Length of new recommendation list is 30
**Output:**
    New recommendation list

---

## 4    Experiments

In this section, we first introduce the dataset for experiments and the experimental setup to evaluate the recommendation performance of our proposed approach. After that, we report the results of standard measures for information retrieval and recommendation (Recall, Precision, and F1 score). Finally, analysis and discussion are presented based on the experimental results.

### 4.1    Dataset

Since there is no existing standard benchmark dataset with annotated term function at the paragraph level for term function-based citation recommendation, we collect and build our data set from ACL Anthology⑤ in the domains of information extraction, sentiment analysis, and recommender systems. In total, 531 research articles with full text and their 2,875 reference articles with title, published year, and abstract are included. As mentioned in section 3.1, all the paragraphs in the related work sections of the 531 research articles have been labeled with one of the four term functions. Since many of the reference articles have been co-cited by the original articles, it is easy for us to figure out their term function distributions.

---

⑤ http://www.aclweb.org/anthology/

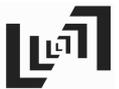

For example, the term function distribution of paper "On the recommending of citations for research papers (McNee et al., 2002)" is (3/8, 4/8, 0, 1/8) in terms of (problem+ method, problem, method+ problem, and method), which can be used as weighting parameters. We use the Python NLTK tool® to perform text pre-processing like removing numbers, stop words, and stemming. Finally, the experiment dataset is stored in a MySQL database.

In our experiment, we randomly partition the dataset into five subsamples and then perform 5-fold cross-validation on the same partition for our approaches and the baseline methods. At each time, four sets were used as training sets for term function weighting, and the remaining one set was used for testing. Also, a small portion of examples split from the training set was used for validation. We performed the process five times and averaged their performance for evaluation.

### 4.2 Experimental setup

- **Baseline models.** We compute the similarity scores between a paragraph and the candidate recommendation articles with the BM25 model (the default values of $b = 0.75$ and $k1 = 1.2$ are applied) and Word2vec based VSM model, respectively. Meanwhile, we restrict the published date of the candidate articles to a year before the original article.
- **Term function weighting-based recommendation models**. In this method, we modify the baseline models by involving term function information and rank the candidate recommendation articles based on the scores computed by Eq. (5).

Moreover, we conduct a comparison study on three fields, including information extraction (300 paragraphs), sentiment analysis (383 paragraphs), and recommender system (510 paragraphs), to analyze the effect of term function on citation recommendation performance in different fields. Therefore, twelve runs in total, including six baseline methods and six term function weighting-based methods. They are BM25 (BM25), word2vec-based VSM (W2V-VSM), BM25 with term function weighting (TFW-BM25), word2vec-based VSM with term function weighting (TFW- W2V-VSM) in information extraction, sentiment analysis, and recommender system, respectively.

### 4.3 Evaluation metrics

Citation recommendation is essentially an information retrieval task. The top-ranked documents are the most important to get the correct recommendation

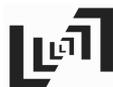

® https://www.nltk.org/

(Wu et al., 2012). Therefore, we employ IR evaluation measures, including Precision, Recall, and F-measure, in our experiments. For a given query (paragraph with its labeled term function) in the test set, we use the original set of references, which were not present while training, as the ground truth (Rokach et al., 2013). In experiments, the number of recommended citations is set as 5, 10, 20, and 30, respectively.

## 4.4 Performance comparison

Table 3, table 4, and table 5 show the result for each compared method on our dataset regarding the three fields, respectively. In the results, the average precision score, recall score, and F1 score in terms of a different number of citations in the recommendation list.

Table 3. Recommendation performance in information extraction.

| Metrics | Runs | Top 5 | Top 10 | Top 20 | Top 30 |
| --- | --- | --- | --- | --- | --- |
| Precision | BM25 | 10.6% | 12.6% | 7.1% | 6.0% |
| | W2V-VSM | 12.8% | 12.0% | 10.8% | 8.9% |
| | TFW-BM25 | 13.6% | 16.6% | 9.6% | 7.4% |
| | TFW-W2V-VSM | **16.4%** | **14.9%** | **13.7%** | **13.0%** |
| Recall | BM25 | 14.9% | 35.1% | 40.0% | 50.6% |
| | W2V-VSM | 17.5% | 38.7% | 48.9% | 58.6% |
| | TFW-BM25 | 19.0% | 46.4% | 53.6% | 61.9% |
| | TFW-W2V-VSM | **23.3%** | **48.7%** | **57.0%** | **66.8%** |
| F1-score | BM25 | 12.4% | 18.5% | 12.1% | 10.7% |
| | W2V-VSM | 14.8% | 18.3% | 17.7% | 17.2% |
| | TFW-BM25 | 17.5% | 24.5% | 16.3% | 13.2% |
| | TFW-W2V-VSM | **19.3%** | **22.4%** | **22.1%** | **21.8%** |

Table 4. Recommendation performance in sentiment analysis.

| Metrics | Runs | Top 5 | Top 10 | Top 20 | Top 30 |
| --- | --- | --- | --- | --- | --- |
| Precision | BM25 | 11.1% | 10.5% | 8.2% | 7.0% |
| | W2V-VSM | 13.6% | 12.9% | 10.5% | 9.5% |
| | TFW-BM25 | 12.9% | 13.0% | 9.1% | 8.2% |
| | TFW-W2V-VSM | **16.7%** | **15.5%** | **13.8%** | **11.0%** |
| Recall | BM25 | 15.6% | 29.6% | 46.3% | 59.3% |
| | W2V-VSM | 19.5% | 34.3% | 52.6% | 67.9% |
| | TFW-BM25 | 18.1% | 36.7% | 51.5% | 69.6% |
| | TFW-W2V-VSM | **27.7%** | **41.0%** | **59.2%** | **73.8%** |
| F1-score | BM25 | 13.0% | 15.5% | 13.9% | 12.5% |
| | W2V-VSM | 16.0% | 18.7% | 17.5% | 16.7% |
| | TFW-BM25 | 15.1% | 19.2% | 15.4% | 17.2% |
| | TFW-W2V-VSM | **20.8%** | **22.5%** | **22.4%** | **19.0%** |

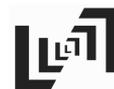

Table 5. Recommendation performance in recommender system.

| Metrics | Runs | Top 5 | Top 10 | Top 20 | Top 30 |
|---|---|---|---|---|---|
| Precision | BM25 | 14.3% | 12.1% | 7.5% | 6.4% |
|  | W2V-VSM | 16.4% | 14.8% | 12.3% | 11.7% |
|  | TFW-BM25 | 17.1% | 15.7% | 10.7% | 8.3% |
|  | TFW-W2V-VSM | **21.7%** | **19.9%** | **16.2%** | **15.2%** |
| Recall | BM25 | 21.3% | 36.2% | 44.7% | 57.4% |
|  | W2V-VSM | 24.7% | 44.4% | 54.2% | 73.1% |
|  | TFW-BM25 | 25.5% | 46.8% | 63.8% | 74.5% |
|  | TFW-W2V-VSM | **26.8%** | **50.1%** | **68.9%** | **77.8%** |
| F1-score | BM25 | 17.1% | 18.1% | 12.8% | 11.5% |
|  | W2V-VSM | 19.7% | 21.9% | 20.0% | 20.2% |
|  | TFW-BM25 | 20.5% | 23.5% | 18.3% | 14.9% |
|  | TFW-W2V-VSM | **24.0%** | **28.5%** | **26.2%** | **25.2%** |

There are some interesting findings from the results:

- First, the term function-based methods outperform the baseline methods. For example, the term function-based BM25 methods improve the baseline BM25 methods by 5.0% (average in the three fields, the same in the following description) on F1 score when the number of recommended citations is set as 20 and improve the baseline by 12.9% on recall when the number of recommended citations is set as 30, while the term function-based Word2vec-VSM methods improve the baseline Word2vec-VSM methods by 5.4% on F1 score when the number of recommended citations is set as 20 and improve the baseline by 9.8% on recall when the number of recommended citations is set as 20. The improvement demonstrates that term function is a useful feature in citation recommendation, particularly in the related work sections.
- Second, there are no significant differences between different fields on both the baseline methods and the term function-based methods. Intuitively, the styles of organizing related studies in the three fields are similar, as proved in Fig. 3. Therefore, it would be interesting to investigate the literature review from different disciplines such as LIS and economy, or different type of publications such as quantitative and qualitative research.
- Third, the Word2vec-VSM model outperforms the BM25 model before and after term function-weighting, indicates the potential of applying term functions to more complex information retrieval such as BERT-based deep learning models to improve both performance and interpretability.

### 4.5 Results analysis and discussion

Our experiment results show that compared to the traditional retrieval models such as BM25 and Word2vec-VSM, which have been proved effective in information retrieval and recommendation, our term function weighting-based strategies recommend a more accurate and structured literature list, demonstrating that term function is an effective feature in citation recommendation, especially when recommending papers for structured literature review. Compared with deep learning-based-recommendation approaches (Zhang et al., 2019), our strategy can generate intuitive explanations of the results for users or system designers, improving system transparency and persuasiveness, trustworthiness, and effectiveness. For example, users can quickly figure out both the content and term function relevance of a recommended item. Not only will this new citation recommendation strategy benefit for semantic scientific information retrieval, but it also benefits automatically structured summarization and literature review generation.

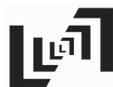

However, there are some limitations to our proposed strategy. In our measurement, we assume that any paper other than the actual citation is not relevant. In fact, there may be multiple papers that provide the same support/evidence and can equally serve as valid citations. Therefore, the above measures probably underestimate the actual performance. It might be interesting to look at which papers other than the one they are citing are relevant. However, that would require subjective and manual relevance judgments. Another is the limited amount of data. Since term function annotation is still a challenging task, it requires field experts and manual work. Although Cheng (2015) is trying to develop automatic term function identification techniques, there is a considerable gap between experiment and application. Therefore, using machine learning models for automatic classification of the term functions will be explored to get more training data. Absolutely, our citation recommendation strategy initiated a research direction of how term function could be used and how to construct a structured literature review system.

## 5  Conclusion and future work

This paper proposed a new citation recommendation strategy based on term functions in the related studies section. Based on the hypothesis that researchers usually organize citations in the related work sections with some patterns, we develop a term function annotation scheme at the paragraph-level. An annotation experiment showed that there were four common patterns of organizing literature in the related work sections. Following this theory, we develop a term function-based citation recommendation framework to recommend users' articles based on their assigned term function of a specific paragraph in the related work sections. Using the "real-world" dataset collected from ACL Anthology, we designed a recommendation experiment in three filed: information extraction, recommender system, and sentiment analysis, with the BM25 model and Word2vec-VSM model as the baseline method. The experiment results show that our proposed citation recommendation strategies outperform all the baseline methods in terms of different evaluation metrics, demonstrating that term function is an effective feature in citation recommendation, especially when recommending papers for structured literature review.

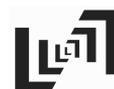

In the future, we will try to develop algorithms to automatically build large-scale data collections with term function for each paragraph. In this way, we can combine the state-of-the-art contextual word embeddings such as SCIBERT (Beltagy, Lo, & Cohan, 2019) with term functions to improve the recommendation performance and provide explainable recommendation results. We will also explore the usefulness of some other features, such as citing time and citation location. More practically, we will implement a recommender system that helps researchers write structured literature reviews more efficiently based on our citation recommendation strategy.

## Acknowledgement

This work is supported by the National Natural Science Foundation of China (Grant No. 7167030644 and 71704137)

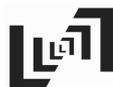

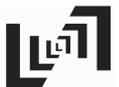

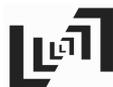